\documentclass[sigconf,natbib]{acmart}

\copyrightyear{2017} 
\acmYear{2017} 
\setcopyright{acmlicensed}
\acmConference{ICTIR '17}{October 1--4, 2017}{Amsterdam, Netherlands}
\acmPrice{15.00}
\acmDOI{10.1145/3121050.3121098}
\acmISBN{978-1-4503-4490-6/17/10}

\hypersetup{
  pdfauthor={Anonymous},
  pdftitle={},
  pdfsubject={},
  pdfkeywords={},
  pdfcreator={LaTeX with hyperref package},
  pdfproducer={pdflatex}
}
\usepackage[inline]{enumitem}
\usepackage[skip=0pt]{caption}
\usepackage{paralist}
\usepackage{xspace}
\usepackage{graphics}
\usepackage[utf8]{inputenc}
\usepackage{epsfig}
\usepackage{multirow}
\usepackage{amsmath,amssymb}
\usepackage{mathtools}
\usepackage{algorithm}
\usepackage{algorithmic}
\usepackage{verbatim}
\usepackage{color}
\usepackage{booktabs}
\usepackage{tabularx}
\usepackage{tikz}
\usepackage{setspace}
\usetikzlibrary{fit}
\usetikzlibrary{positioning}

\newcounter{todocnt}

\newcounter{latercnt}

\usepackage{acronym}
\acrodef{RL}{Reinforcement Learning}
\acrodef{IRL}{Inverse Reinforcement Learning}
\acrodef{SERP}{search engine result page}
\acrodef{IR}{Information Retrieval}
\acrodef{MDP}{Markov Decision Process}

\def\:{\hskip0pt} % \: for hyphenation xxx\:---\:xxx  xx-\:xxx

\newcommand{\is}{interactive system\xspace}

\newcommand{\descuf}{static user features\xspace}
\newcommand{\dinuf}{dynamic user features\xspace}

\title{Towards Learning Reward Functions from User Interactions}
\author{Ziming Li}
\orcid{}
\affiliation{%
\institution{University of Amsterdam}
\city{Amsterdam}
\country{The Netherlands}
}
\email{z.li@uva.nl}

\author{Julia Kiseleva}
\orcid{}
\affiliation{%
\institution{UserSat.com \& University of Amsterdam}
\city{Amsterdam}
\country{The Netherlands}
}
\email{j.kiseleva@uva.nl}

\author{Maarten de Rijke}
\orcid{0000-0002-1086-0202}
\affiliation{%
\institution{University of Amsterdam}
\city{Amsterdam}
\country{The Netherlands}
}
\email{derijke@uva.nl}

\author{Artem Grotov}
\orcid{}
\affiliation{%
\institution{University of Amsterdam}
\city{Amsterdam}
\country{The Netherlands}
}
\email{a.grotov@uva.nl}

\begin{document}

\begin{abstract}
     % Let's do a Kent Beck style abstract in 4 sentences:
     %\todo[1. State the problem]
  
     In the physical world, people have dynamic preferences, e.g., the same situation can lead to satisfaction for some humans and to frustration for others. Personalization is called for. The same observation holds for online behavior with interactive systems.
     It is natural to represent the behavior of users who are engaging with interactive systems such as a search engine or a recommender system, as a sequence of actions where each next action depends on the current situation and the \emph{user reward} of taking a particular action.
     By and large, current online evaluation metrics for interactive systems such as search engines or recommender systems, are static and do not reflect differences in user behavior. They rarely capture or model the \emph{reward} experienced by a user while interacting with an interactive system. We argue that knowing a user's reward function is essential for an interactive system as both for learning and evaluation.
     We propose to learn users' reward functions directly from observed interaction traces. In particular, we present how users' reward functions can be uncovered directly using \emph{inverse reinforcement learning} techniques. We also show how to incorporate user features into the learning process.
     Our main contribution is a novel and dynamic approach to restore a user's reward function. We present an analytic approach to this problem and complement it with initial experiments using the interaction logs of a cultural heritage institution that demonstrate the feasibility of the approach by uncovering different reward functions for different user groups.
 
\end{abstract}




\maketitle
 
\newcommand{\rqmain}{The main aim of this paper is {\emph{to study how user reward functions can be learned from their interactions}}, which we break down into the following concrete research questions.}

\newcommand{\rqone}{\textbf{RQ1:} {How to define user reward?}\xspace}
\newcommand{\rqtwo}{\textbf{RQ2:} {How to recover a user reward function?}\xspace}
\newcommand{\rqthree}{\textbf{RQ3:} {How to incorporate user features into the learning process?}\xspace}

\newcommand{\rqmainconc} {\emph{how reward functions can be directly learned from users' interplays with an interactive system}}

\newcommand{\negskip}{\vspace*{-.5\baselineskip}}

\section{Introduction}
\label{foursec:intro}

Understanding and modeling user behavior is a fundamental problem for any interactive system as insight into user behavior will lead towards ``proper'' evaluation: what satisfies user needs and what frustrates users. We know that users have different preferences and can display different behavior~\cite{wei_wsdm_2017,Kosinski_2013}. Despite this key lesson, the evaluation metrics in use today do not take differences in user behavior into account. Existing methods are directed towards generalization~\citep{hofmann2011_ecir} rather than personalization~\citep{zhang_aaai_2016}. This stops us from deciphering fine-grained user feedback. The idea of designing an \is that dynamically reacts to user actions by employing the reinforcement learning (RL) paradigm as proposed in Figure~\ref{fig:rl} is appealing. A key problem preventing this is that we do not know the users' \emph{true} reward functions.

\tikzset{
  pobl/.style={
    inner sep=0pt, outer sep=0pt, fill=#1,
  },
  pobl gron/.style n args={2}{
    pobl=#1, rounded corners=#2,
  },
  pics/person/.style n args={3}{
    code={
      \node (-corff) [pobl=#1, minimum width=.25*#2, minimum height=.375*#2, rotate=#3, pic actions] {};
      \node (-pen) [minimum width=.3*#2, circle, pobl=#1, outer sep=.01*#2, anchor=south, rotate=#3, pic actions] at (-corff.north) {};
      \node (-coes dde) [pobl gron={#1}{1pt}, anchor=north west, minimum width=.12125*#2, minimum height=.25*#2, rotate=#3, pic actions] at (-corff.south west) {};
      \node [pobl=#1, anchor=north, minimum width=.12125*#2, minimum height=.15*#2, rotate=#3, pic actions] at (-coes dde.north) {};
      \node (-coes chwith) [pobl gron={#1}{1pt}, anchor=north east, minimum width=.12125*#2, minimum height=.25*#2, rotate=#3, pic actions] at (-corff.south east) {};
      \node [pobl=#1, anchor=north, minimum width=.12125*#2, minimum height=.15*#2, rotate=#3, pic actions] at (-coes chwith.north) {};
      \node (-braich dde) [pobl gron={#1}{.75pt}, minimum width=.075*#2, minimum height=.325*#2, outer sep=.0064*#2, anchor=north west, rotate=#3, pic actions] at (-corff.north east)  {};
      \node [pobl=#1, minimum width=.05*#2, minimum height=.2*#2, outer sep=.0064*#2, anchor=north west, rotate=#3, pic actions] at (-corff.north east) {};
      \node (-braich chwith) [pobl gron={#1}{.75pt}, minimum width=.075*#2, minimum height=.325*#2, outer sep=.0064*#2, anchor=north east, rotate=#3, pic actions] at (-corff.north west) {};
      \node [pobl=#1, minimum width=.0375*#2, minimum height=.2*#2, outer sep=.0064*#2, anchor=north east, rotate=#3, pic actions] at (-corff.north west) {};
      \node (-fit person) [fit={(-pen.north) (-braich dde.east) (-coes chwith.south) (-braich chwith.west)}] {};
      \node (-pwy) [below=25pt of -fit person, every pin] {\tikzpictext};
      \draw [every pin edge] (-fit person) -- (-pwy);
    },
  },
}

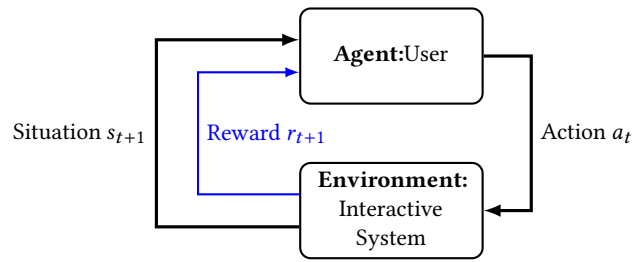
\begin{figure}[h]
\centering
\tikzstyle{block} = [rectangle, draw, text width=7em, text centered, rounded corners, minimum height=4em]    
\tikzstyle{line} = [draw, -latex]
\tikzstyle{line_thick} = [draw, -latex, line width=1.2pt]

\begin{tikzpicture}[node distance = 6.5em, auto, thick]
  [
    every pin edge/.append style={latex-, shorten <=-2.5pt},
  ]
    %\node [block](Agent) {};
    %\draw pic (person) [pic text={\textbf{Agent}:User}]
    %{person={pink}{22pt}{0pt}};
    \node [block](Agent) {\textbf{Agent:}User};
    %\draw pic[pic text={\textbf{Agent}:User}]
    \node [block, below of=Agent] (Environment) {\textbf{Environment:} Interactive System};
    \path [line_thick] (Agent.0) --++ (2em,0em) |- node [near start]{Action $a_t$} (Environment.0);
    \path [line_thick] (Environment.190) --++ (-6em,0em) |- node [near start] {Situation  $s_{t+1}$} (Agent.170);
    \path [blue] [line] (Environment.170) --++ (-4.25em,0em) |- node [near start, right] {Reward $r_{t+1}$} (Agent.190);
\end{tikzpicture}
\smallskip
\caption{The user-system interface.}
\vspace{-4ex}
\label{fig:rl}
\end{figure}
 
Let's consider an example of interactive system---a web search engine. Assume a user issues the query ``panda'' and the search engine returns a diverse \ac{SERP} that contains answers from various verticals: text, images, videos. If our user is a child, he will most likely click on an image result, and adults may prefer to read a Wikipedia page. According to existing evaluation paradigms, based, e.g., on the number of satisfied clicks~\citep{luo2015session}, both outputs are successful because the user clicks on the \ac{SERP}~\citep{borisov_www_2016}. 
%In case of NDCG~\citep{hofmann2011_ecir}, we expect to a best answer not taking into account user preferences. 
To increase user satisfaction, search engines are currently showing diverse \acp{SERP}, but this is no silver bullet. With the popularity of mobile phones, we are moving to an era of personal assistants on mobile devices~\citep{kiseleva_sigir_2016, kiseleva_chiir_2016} and direct answers~\citep{Williams_www_2016}, where the screen size is small or a system is expected to give the best result that directly answers user needs. In such scenarios it is not an option for an \is to offer a broad selection of alternatives. Instead, the \is has to discover, during successive interactions, a user's preferences.

Our long-term goal is to dynamically process user feedback for evaluating user reward during an interactive session and to respond accordingly as presented in Figure~\ref{fig:rl}. For example, if a user clicks on images after issuing the query ``panda,'' an \is can infer a user's reward to provide a \emph{better} experience for the next step in the interaction.

\rqmain

\negskip
\paragraph{\rqone}
We model the interactive user-system interface in Figure~\ref{fig:rl} using \ac{RL}~\cite{sutton_irl_2012}.
More specifically, a user is an agent who interacts with an environment, which is an \is, in a sequential manner with discrete time steps, $t = 0, 1, 2, 3,\dots$.
At each time step $t$, the user receives some representation of the system, which we call \emph{situation}, $s_t \in S$, where $S$ is the set of
possible system situations, and on that the user performs an action, $a_t \in A$, where $A$ is the set of actions possible in the situation $s_t$. 
We propose the following way of modeling the user reward function: one time step later, 
the user receives a numerical reward, $R_{t+1} \in R$, and finds himself in a new situation, $s_{t+1}$. 
While examining interactive user behavior, we should consider the reward function unknown and to be restored though empirical detection. This is specifically important for multi-attribute reward functions as in our case.
A unit of user interactions with the system within some time period $[t_i,t_{i+j}]$, called a \emph{user session}.
  
\negskip
\paragraph{\rqtwo}
Our problem fits the setting of \ac{IRL}~\cite{ng_icml_2000}, which is defined as follows: \textbf{given:} \begin{inparaenum}[(1)] \item sequential users' interactions with a system over time in a variety of circumstances, and \item a system model; \textbf{determine:} the reward function of the users.\end{inparaenum}

\negskip
\paragraph{\rqthree}
\ac{IRL} techniques have successfully been applied for the apprenticeship (or imitation) learning problem~\citep{ng_icml_2000}, to discover the reward function whose optimization would produce \emph{desirable behavior}. One of the prominent examples is self-driving cars, where one recovers the reward function based only on expert driving behavior. In contrast, we attempt to restore reward function(s) covering various types of user behavior (similar to driving styles in~\cite{abbeel_icml_2004}). 
Similar to~\cite{banovic_chi_2016} we incorporate into our learning process user features that can be organized in two groups: \begin{inparaenum}[(1)] \item \emph{\descuf} remain unchanged during the user session, e.g., age, gender; \item \emph{\dinuf} describe user behavior in the particular situation $s_t$, e.g., time spent.\end{inparaenum}\ 

To confirm our hypothesis that different types of users have different reward functions, we perform a preliminary reward learning experiment for which we choose onsite logs of physical interactions in a museum because user features are explicitly given.
\if0
\medskip\noindent%
The remainder of this paper is organized as follows. Sec.~\ref{sec:rel_work} describes earlier work and background. 
Sec.~\ref{sec:framework} describes the suggested approach to define and to recover user reward functions.
We present our preliminary experimental results in Sec.~\ref{sec:experiments}. Finally, we conclude and identify future directions in Sec.~\ref{sec:conc_fw}.
\fi

\vspace{-1ex}  

\section{Background and Related Work}
\label{sec:rel_work}

\paragraph{Evaluation}
\ac{IR} is about getting the right information to the right people in the right way. Evaluation has historically been one of \ac{IR}'s key concerns. Offline, system-oriented evaluation, with a strong focus on assessing the degree to which a system is able to successfully identify documents that are relevant to a query, has received considerable attention~\citep{sanderson-test-2010}. In parallel, user-oriented evaluation methods for interactive information retrieval received considerable attention~\citep{kelly-methods-2009}. Increasingly, though, there is a realization that system aspects and user aspects should be assessed in tandem. In online experimentation the two aspects naturally come together~\citep{hofmann-online-2016}.

Online controlled experiments, such as A/B testing or interleaving, have become widely used techniques for controlling and improving search quality based on data-driven decisions~\citep{kohavi_kdd_2014}. This methodology has been adopted widely~\citep{Deng_www_2014,Tang_kdd_2010,Bakshy_kdd_2013,drutsa_www_2015}.  An A/B test is a between-subject test designed to compare two variants of a method (e.g., ranking on the \ac{SERP}, ad ranking, colors and fonts of the web result title) at the same time by exposing them to two user groups and by measuring the difference between them in terms of a \emph{key metric} (e.g., revenue, number of visits, etc.).

There are many existing studies towards better online evaluation that are devoted to improving the sensitivity of our measurement methods~\citep{schuth-multileaved-2014}, inventing new metrics~\citep{Drutsa_wsdm_2015,Dupret_wsdm_2013} or improving existing ones~\citep{drutsa_www_2015}. An important goal of recent studies is to make metrics more consistent with long-term goals~\citep{kohavi_kdd_2014}.
User engagement metrics show different aspects of user experience. For instance, they can reflect \begin{enumerate*} \item \emph{user loyalty} -- the number of sessions per user~\cite{Song_www_2013}; \item \emph{user activity} -- the number of visited web pages~\citep{Lehmann_cikm_2013} or the absence time~\citep{Dupret_wsdm_2013}.\end{enumerate*} Periodicity engagement metrics of user behavior, which result from the discrete Fourier transform of state-of-the-art engagement measures, have also been proposed~\citep{Drutsa_wsdm_2015}. 
Few studies have looked at evaluating intelligent assistants in online settings~\citep{kiseleva_sigir_2016, kiseleva_chiir_2016, Williams_www_2016, Williams_sigir_2016}, where user satisfaction~\citep{kelly-methods-2009} is defined and predicted at the session level.

Most existing studies are directed towards generalization from user interaction rather than understanding the behavior of individuals. 
Very few works~\cite{azzopardi_sigir_2014} have explored why users behave in particular ways by applying economic models.

\negskip
\paragraph{Reinforcement learning in interactive systems}
Several authors have adopted a \acf{RL} perspective on \ac{IR} problems. \citet{hofmann2011_ecir,hofmann-balancing-2013} seem to have been the first; they use \ac{RL} for online evaluation and online learning to rank and define reward functions directly in terms of NDCG~\citep{hofmann2013_wsdm}. Later work on \ac{RL} in \ac{IR} predefined reward functions as the number of satisfied clicks in session search~\citep{luo2015learning, luo2015session}. \citet{odijk-dynamic-2015} use \ac{RL} for query modeling and define reward in terms of retrieval performance (NDCG). 
Applications of \ac{IRL} in interactive systems are relatively rare. \citet{ziebart_iui_2012} use \ac{IRL} for predicting the desired target of a partial pointing motion in graphical user interfaces. \citet{monfort_aaai_2015} use \ac{IRL} to predict human motion when interacting with environment. It is not straightforward to apply \ac{IRL} for interactive systems as it is often unclear how one should deal with user features~\citep{banovic_chi_2016}.  

\smallskip\noindent%
To summarize, the key distinctions of our work compared to previous work are that we introduce a new problem (recovering user reward functions from interaction data) and propose a method to address the problem.

\vspace{-1ex}

\section{Learning User Reward Functions}
\label{sec:framework}

We start by investigating~\emph{\rqone}
To model the user reward function, presented in Figure~\ref{fig:rl}, we use a finite \ac{MDP}.
An \ac{MDP} is a tuple $(\mathbf{S}, \mathbf{A},$ $T, d_0,\gamma, R)$, where $\mathbf{S}$ is a set of $N$ states (possible system situations);
$\mathbf{A}$ is a set of $K$ actions;
$T$ is a set of state transition probabilities $P_{sa}(\cdot)$ is a state transition probability upon taking action $a$ in state $s$;
the initial distribution of states is $d_0$ and $\gamma \in [0,1)$ is a discount factor;
$R$ is a user reward function, where $R(s,a)$ is a reward given for action $a$ in situation $s$. 
Given a current state $s$ and action $a$ together with any next state $s'$, the expected value of the next reward is: $r(s,a,s')$ = $\mathbb{E}(R_{t+1}\mid S_t=s, A_t=a, S_{t+1}=s')$.% For simplicity, we write $R(s)$ instead of $R(s,a)$.

An MDP without reward function is denoted as MDP$/$R, i.e., a tuple $(\mathbf{S}, \mathbf{A},$ $T, d_0,\gamma)$. 
Let $\mathbf{f} : S \rightarrow [0,1]^k$ be a vector of features over states. There is a ``true'' reward function $R$ that is given by a linear combination of $k$ features $f_i$ with weights $\theta_i$ where $\theta^T \in \mathop{\mathbb{R}}^k$.
We assume that the reward function is defined as $R(s)=\theta^T \cdot \mathbf{f}(s)$. In our setting, $\mathbf{f}$ is a vector of features describing user behavior, e.g., time a user spent interacting with the system state. The desired vector $\theta^T$ specifies the relative weighting between these features.

A \emph{policy} is a map $\pi$ from situations, $s_t \in S$, and actions, $a_t \in A$, to the probability $\pi(a_t\mid s_t)$ of taking action $a_t$ when in state $s_t$. The value function $V$ for policy $\pi$ is:
\begin{equation}
  \mbox{}\hspace*{-.2cm}
  \mathbb{E}_{s_0 \sim d_0}[V^\pi (s_0)] =  \mathbb{E}\left[\sum^{\infty}_{t=0} \gamma^t R(s_t)|\pi] 
  %=\mathop{\mathbb{E}}[\sum^{\infty}_{t=0} \gamma^t \theta \mathbf{f}(s_t)\mid \pi]
  =\theta^T \cdot \mathbb{E}[\sum^{\infty}_{t=0} \gamma^t \mathbf{f}(s_t)\mid \pi\right].
\end{equation}
The expectation is taken with respect to random sequences of situations $s_0,s_1,\dots$ drawn from the starting situation $s_0 \sim d_0$.
The goal of \ac{RL} is to find $\pi$ such that $V^\pi(s)$ is maximized. There exists at least one optimal policy $\pi^*$ such that $V^\pi(s)$ is simultaneously maximized for all $s_t \in S$ by $\pi^*=\pi$~\citep{ng_icml_2000}. 

We need to estimate a user's features expectations $\mu_{E}$. Given a set of $m$ user sessions $\{s^{i}_0, s^{i}_1, s^{i}_2, \dots \}_{i=0}^m$, generated by users, we empirically estimate $\mu_{E}$ as 
$\hat{\mu}_{E} = 1/m \cdot \sum_{i=0}^m \sum_{t=0}^\infty \gamma^t\mathbf{f}(s^{i})$.

Next, we consider \emph{\rqtwo}
The problem of \ac{IRL} is to find a reward function that can explain the observed user behavior. We formulate the problem as follows:
\textbf{given} 
\begin{inparaenum}[(1)] 
\item  an MDP$/$R,
\item  $\mathbf{f}$,
\item user feature expectations $\mu_{E}$;
\end{inparaenum}
\textbf{determine} a policy whose performance is close to the observed user group behavior based on the unknown user reward function $R=\theta^T\cdot\mathbf{f}$.
There are a number of available \ac{IRL} methods~\cite{boularias2011relative,ziebart_tit_2013, ng_icml_2000, abbeel_icml_2004,ziebart_aaai_2008}. For our preliminary experiments we adopt Maximum Entropy \ac{IRL} (MaxEnt)~\citep{ziebart_aaai_2008} to recover the user reward functions.
\vspace{-1ex}

% Table with results moved earlier to ensure it's on the page of the Experimental section...

\section{Incorporating User Features}
\label{sec:experiments}

We report on a preliminary experiment aimed at assessing the feasibility of uncovering the reward function from interaction data.

\paragraph{Data}
The dataset we use~\citep{hashemi2017skip} is extracted from the physical interaction logs of an archaeological museum. Besides common exhibitions, this museum also provides additional information that can be obtained from different POIs. The contents at each POI are based on one specific topic and there are 8 topics in total. According to the corresponding topic, each POI shows 3 related objects and the objects at different POIs can be accessed in any order. Users can enter their personal information and preference at the beginning to personalize the contents being shown. 
%But for this dataset, the personalizing process has no affect on our IRL model. 
The original dataset consists of 5 months of onsite logs with about 21,000 sessions. Each record contains one user's personal information (e.g., age and language) and the interaction order with different objects at POIs. The starting time and how long the interaction lasts for each object are also recorded. After filtering the sessions which did not have any interactions or necessary user information (such as age), 4,694 out of 21,000 interaction sessions remain and constitute our final data.
\negskip
\paragraph{Experimental design}
To address \emph{\rqthree} we propose the following experiment.
We focus on exploring the difference between the reward functions of different groups. According to the user's age, we divide the data into two groups, child and adult, with 1,135 and 3,559 sessions, respectively. To indicate each interaction situation, we consider three kinds of features: ``\emph{topic}'' (8 different types, including: \emph{appearance}, \emph{death}, \emph{religion}, \emph{architecture}, \emph{entertainment}, \emph{food}, \emph{trade}, \emph{army}), ``\emph{object order}'' (3 objects for each topic) and ``\emph{duration time}'' (discretized in 3 bins denoting 0--30s, 30--90s and more than 90s). One-hot encoding is used in our experiment as some features are categorical. In this manner, 14 features are selected and 72 situations are defined. With respect to the action feature, we use the object's topic number to denote an action; taking a specific action means the user will transition to the situations that has the same topic number. We identified 8 actions in this dataset. In terms of transition probability, we simply count the occurrence frequency in the behavior history to estimate the probability for possible situations when the current situation and action are determined. 

\paragraph{Experiments and results}

\begin{table*}
\caption{The recovered weights of reward function.}
\resizebox{\linewidth}{!}{
\begin{tabular}{l*{11}{c}}
\toprule
\textbf{Group} & \textbf{Appearance} & \textbf{death} & \textbf{religion}  & \textbf{architecture}  & \textbf{entertainment}  & \textbf{food} & \textbf{trade} & \textbf{army} & \textbf{object1} & \textbf{object2} & \textbf{object3} \\
\midrule
Adult   & 0.3378  & 0.1340  & 1.0539  & -0.7106  & 0.9119  & 1.3577  & 0.4174  & 0.6783  & 0.6161  & 0.6556  & 0.6171  \\
\midrule
Child   & 0.9436  & 0.6479  & 1.2215  & 0.2658   & 1.2394  & 1.5001  & 0.7700  & 0.4649  & 0.9405  & 0.7115  & 0.7368 \\
\bottomrule
\end{tabular}
}
\vspace{-3ex}
\label{table:1}
\end{table*}

\negskip
Table~\ref{table:1} shows part of the learned weights of the reward function.  
The feature ``\emph{architecture}'' negatively contributes to the user reward function for the Adult group. With respect to the feature ``\emph{death},'' the Child group has a higher weight, which makes intuitive sense as children are more curious about scary contents. For the Adult group, objects of ``\emph{food}'' contribute most to the reward while it is also the most popular topic for children. For the Adult group, ``\emph{religion}'' has a higher weight compared to ``\emph{appearance}'' while it is same for the Child group. But for ``\emph{trade}'' and ``\emph{army}'', these two groups have different preference. Another interesting phenomena is that most weights of object order of the Adult group are in a small scope compared to Child group, which can be explained as that adults will interact with all the three objects with the same topic while most children will only view objects in the front of the object order. 
As we can see, there really does exist a difference between different groups' reward functions and this kind of difference was also reported in the setting of self-driving cars~\citep{abbeel_icml_2004}.

\vspace{-1ex}

\section{Conclusions and Future Work}
\label{sec:conc_fw}

In this conceptual paper we investigated \rqmainconc.

First, we explored~\emph{\rqone} We used a container of different features (such as system features and user features) to represent all possible situations. The user reward function, then, is a linear combination of situation features that we used to explain demonstrated user behavior. 

Second, to answer~\emph{\rqtwo} we proposed to use \ac{IRL} techniques. We adopted maximum entropy \emph{IRL} to recover the users reward functions. Our experimentation with a physical interaction dataset showed that the reward functions of different user groups have different priorities about features. Some features have a bigger impact on one user group's reward than on another group's. 

Third, we studied~\emph{\rqthree}
In the dataset used, explicit user features are given. We see two ways of incorporating user features into the learning process: \begin{inparaenum}[(1)] \item Grouping based on \descuf; explicit user features are needed and we recover the reward functions for all the groups separately; the drawback is that groups need to be predefined and the number of user features should be relatively small.  \item Using user features to describe the situation.\end{inparaenum}\ In many scenarios, such as web search, explicit user features may be unavailable and the first method is not applicable. How to adjust the reward for different users without predefined groups? 
\if0
A possible direction is to select features that can identify behavioral differences between users and add these features to the feature container of each state. 
\fi

In conclusion, recovering user reward functions is \begin{inparaenum}[(1)] \item feasible, \item a promising direction, and \item applicable in many scenarios, including personal assistants, web search, recommender system.\end{inparaenum}

As we are only at the beginning of our investigations into user reward functions, many questions remain open. \begin{inparaenum}[(1)] \item How can we make the reward function more complex, e.g., non-linear, rather than assuming that the function is a linear combination of situation features? \item How can we make the system learn the rewards for different users automatically and return personalized rewards? To achieve this goal, user features should be taken into account during the learning process which can balance the reward for different users. Besides \descuf, \dinuf\ could also influence the decisions of users and need to be considered if possible.  \item How can we solve the computational problem of very large state spaces? The presence of very many features implies that many states will be defined, which in turn may reduce learning efficiency. Automatic feature construction and feature selection can be considered.  \item With more diverse and complicated features being considered, how can we adopt emerging techniques (such as deep inverse reinforcement learning) to mine user interaction scenarios effectively and efficiently?\end{inparaenum}

% requires \usepackage{setspace}

\begin{spacing}{1}
\smallskip\noindent\small
\textbf{Acknowledgments.}
This research was supported by
Ahold Delhaize,
Amsterdam Data Science,
the Bloomberg Research Grant program,
the Criteo Faculty Research Award program,
the Dutch national program COMMIT,
Elsevier,
the European Community's Seventh Framework Programme (FP7/2007-2013) under
grant agreement nr 312827 (VOX-Pol),
the Microsoft Research Ph.D.\ program,
the Netherlands Institute for Sound and Vision,
the Netherlands Organisation for Scientific Research (NWO)
under pro\-ject nrs
612.\-001.\-116, % ImFIRE
HOR-11-10, % HORIZON
CI-14-25, % MediaNow
652.\-002.\-001, % Re-Search
612.\-001.\-551, % CLEAR
652.\-001.\-003, % NEWEL
and
Yandex.
All content represents the opinion of the authors, which is not necessarily shared or endorsed by their respective employers and/or sponsors.
\end{spacing}

\negskip
\bibliographystyle{abbrvnat-tweaked}
\bibliography{bibliography}

\end{document}